# Building Defect Prediction Models by Online Learning Considering Defect Overlooking

Nikolay FEDOROV†, Yuta YAMASAKI††, Masateru TSUNODA††,
Akito MONDEN†, Amjed TAHIR†††, Kwabena Ebo BENNIN††††,
Koji TODA†††††, and Keitaro NAKASAI††††††

**SUMMARY** Building defect prediction models based on online learning can enhance prediction accuracy. It continuously rebuilds a new prediction model, when a new data point is added. However, a module predicted as "non-defective" can result in fewer test cases for such modules. Thus, a defective module can be overlooked during testing. The erroneous test results are used as learning data by online learning, which could negatively affect prediction accuracy. To suppress the negative influence, we propose to apply a method that fixes the prediction as positive during the initial stage of online learning. Additionally, we improved the method to consider the probability of the overlooking. In our experiment, we demonstrate this negative influence on prediction accuracy, and the effectiveness of our approach. The results show that our approach did not negatively affect AUC but significantly improved recall.

*keywords:* Fault prediction, CVDP, overlooking, false negative, oversight prevention

## 1. Introduction

Software testing is one of the key activities to find defects. However, due to scarce resource availability and software development duration, testing can be limited to few number of modules [8]. Defect prediction is one of the major approaches to suppressing remaining defects. When a module is regarded as defective by the prediction model, it is tested thoroughly (i.e., more effort is spent on testing it). In contrast, a module regarded as non-defective is tested much more lightly [4]. When the accuracy of the prediction model is high, both low testing costs and high software quality can be achieved.

To build a defect prediction model, training data based on the previous version history is often used. For instance, during the development of version 1.0, data such as the number of found defects and the complexity of the modules are recorded. Next, using this data, a defect prediction model for the next version (e.g., 1.1) is built. Lastly, during the development of version 1.1 (i.e., on test data), defects of each module are predicted using the prediction model built in the previous stage. The procedure is called cross-version defect prediction (CVDP).

However, the accuracy of CVDP is often low. This is because when the version is different between learning and test data, effective independent variables of the prediction model are often different. This is regarded as an external validity issue of defect prediction [1]. To address the problem, online learning approaches have been proposed recently [6]. When a new data point is added, online learning adds it to the learning dataset and rebuilds a new prediction model. Using this approach, results of software testing is collected and utilized to enhance prediction accuracy during development.

Fig. 1 illustrates an example of defect prediction by online learning. Each module is tested sequentially from the top to the bottom. After module t9 is tested (i.e., before t5), a prediction model M1 is built. The learning dataset includes modules t1 and t9, where an independent variable is the lines of code (LOC), and a dependent variable is the test result. The test result of t5 is predicted by M1. After t5 is tested, model M2 is built based on the data of t1, t9, and t5. The test result of t7 is predicted by M2.

However, the learning dataset is not always correct, due to defect overlooking during software testing. To the best of our knowledge, past studies have not considered the influence of defect overlooking on the performance of defect prediction via online learning. The major contributions of the paper is that we assess and demonstrate the influence of defect overlooking on online learning, and propose a new method to suppress the influence.

† The authors are from Okayama University, Japan.
†† The authors are with Kindai University, Japan.
††† The author is with Massey University, New Zealand.
†††† The author is with Wageningen University & Research, the Netherlands.
††††† The author is with Fukuoka Institute of Technology, Japan.
†††††† The author is with Osaka Metropolitan University College of Technology.

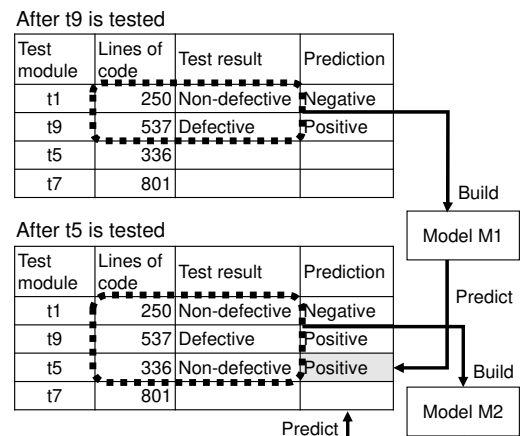

**Fig. 1** Example of defect prediction by online learning

**Fig. 2** Example of two types of defect overlooking

**Fig. 3** Example of fixed prediction method proposed in study [7]

**Fig. 4** Example of proposed method which quits fixed prediction.

## 2. Defect Overlooking

There are two types of defect overlooking. The overlooking is pointed out in study [7] (Note that the study did not use online learning). Both overlooking makes learning dataset used by online learning incorrect, as explained below. This section explains the two types of overlooking.

**Type 1 overlooking**: When a defect prediction model predicts a negative result (i.e., "non-defective"), developers will typically write fewer test cases for those modules [4] in order to efficiently allocate testing resources [5][8]. As a result, defects could be overlooked by the test, and the module might be regarded as "non-defective," even if the module is defective. We call this case as *Type 1 overlooking*. This means that overlooking of defects occurs due to fewer test cases based on negative prediction.

Type 1 overlooking could negatively affect the accuracy of prediction models produced by online learning. In Figure 2, the column "test result" considers only defects during testing, while "actual result after testing" also considers defects after testing was done (e.g., when the software is released). In the example, we assume that defects are overlooked when the prediction is negative due to fewer test cases on a certain probability $n\%$. That is, when the "Prediction" column is "Negative" in Fig. 2, the "Test result" column is "Non-defective" with $n\%$ probability. The probability depends on how much testing resource is assigned (i.e., test cases are made) to negative prediction modules.

Module t1 is regarded as non-defective based on the test outcomes, and they are used as learning data for model M2. However, based on the actual result, the learning data is incorrect, and it should be set as defective. As a result, the accuracy of model M2 becomes low, and the model predicts module t7 as "non-defective" erroneously.

Note that the influence of Type 1 overlooking to prediction accuracy is not only affected by the probability $n$. For instance, when a model accurately predicts non-defective modules as negative, Type 1 errors seldom occur even if $n$ is high. Additionally, when most modules are predicted as positive, Type 1 errors seldom occur.

**Type 2 overlooking**: Even when the prediction is positive (i.e., "defective"), and many test cases are applied, defects are sometimes overlooked during testing. We call this case as defect overlooking by positive prediction. This could occur even when testing resources are not allocated by defect prediction. Modules t5 in the figure is an example of such case. This is because we cannot find all defects perfectly by testing. Based on large-scale data from cross-companies [2], about 17% of defects are overlooked during integration testing.

## 3. Handling Defect Overlooking

**Fixed prediction**: To suppress the influence of Type 1 overlooking, we propose to apply the fixed prediction method proposed in study [7], which turns negative into positive prediction on $m$ negative-predicted modules (i.e., during the early iteration of online learning). The value $m$ is 10% of the number of all test target modules. For instance, when the number of test target modules is 100, $m$ is set as 10. As shown in the column "Fixed prediction" of Fig. 3, the fixed prediction of t1 and t7 module is set as "positive". This restrains Type 1 overlooking and keeps test results (i.e., learning data) accurate with high probability. This is because when the prediction is "positive", many more test cases are applied to the module.

Note that the method uses the model's prediction when $m$ negative-predicted modules have been tested. We could not avoid Type 2 overlooking (module t5 on the figure) by the method. The study [7] did not rebuild prediction models by online learning. Hence, the effect of the method to online learning is unclear.

**Table 1.** Accuracy of reference models

| Dataset | AUC | Precision | Recall | F1 score |
|---------|------|-----------|--------|----------|
| Ant     | 0.74 | 46.3%     | 72.0%  | 56.3%    |
| Prop    | 0.63 | 17.8%     | 55.2%  | 26.9%    |
| Synapse | 0.70 | 56.1%     | 65.8%  | 60.5%    |

**Table 2.** Difference between reference and ordinary models

| Type 1 overlooking | Ant | | | | Prop | | | | Synapse | | | |
|---|---|---|---|---|---|---|---|---|---|---|---|---|
| | AUC | Precision | Recall | F1 score | AUC | Precision | Recall | F1 score | AUC | Precision | Recall | F1 score |
| 20% | -0.01 | 0.3% | -3.4% | -0.9% | 0.00 | 0.5% | -2.6% | 0.2% | -0.02 | -0.6% | -4.3% | -2.2% |
| 40% | -0.01 | 0.9% | -5.2% | -1.1% | -0.01 | 0.5% | -6.1% | -0.3% | -0.03 | -0.2% | -9.2% | -4.4% |
| 60% | -0.02 | 4.9% | -11.8% | -1.3% | -0.03 | 0.3% | -12.5% | -1.7% | -0.05 | -0.5% | -15.8% | -8.5% |
| 80% | -0.05 | 7.5% | -20.6% | -4.0% | -0.06 | 0.2% | -23.9% | -4.6% | -0.08 | 1.0% | -28.7% | -16.5% |
| 100% | -0.22 | -32.3% | -65.7% | -19.5% | -0.13 | -13.3% | -53.4% | -12.7% | -0.17 | -29.6% | -56.6% | -33.4% |

**Table 3.** Difference between ordinary and fixed prediction models

| Type 1 overlooking | Ant | | | | Prop | | | | Synapse | | | |
|---|---|---|---|---|---|---|---|---|---|---|---|---|
| | AUC | Precision | Recall | F1 score | AUC | Precision | Recall | F1 score | AUC | Precision | Recall | F1 score |
| 20% | -0.03 | -8.0% | 6.6% | -4.5% | -0.02 | -3.3% | 7.8% | -3.1% | -0.02 | -6.8% | 7.3% | -1.3% |
| 40% | -0.02 | -8.4% | 6.9% | -4.4% | -0.02 | -3.3% | 8.0% | -2.8% | -0.02 | -7.6% | 9.4% | -0.4% |
| 60% | -0.02 | -12.0% | 12.2% | -4.2% | 0.00 | -3.0% | 11.7% | -1.6% | 0.00 | -6.9% | 13.7% | 3.0% |
| 80% | 0.00 | -13.6% | 17.7% | -1.5% | 0.02 | -3.1% | 18.8% | 0.6% | 0.03 | -8.2% | 23.5% | 9.9% |
| 100% | 0.17 | 26.9% | 59.5% | 13.5% | 0.08 | 10.3% | 40.3% | 7.5% | 0.11 | 22.0% | 48.1% | 25.2% |

**Table 4.** Difference between ordinary and proposed models

| Type 1 overlooking | Ant | | | | Prop | | | | Synapse | | | |
|---|---|---|---|---|---|---|---|---|---|---|---|---|
| | AUC | Precision | Recall | F1 score | AUC | Precision | Recall | F1 score | AUC | Precision | Recall | F1 score |
| 20% | -0.01 | -4.3% | 4.7% | -1.9% | -0.01 | -2.0% | 4.9% | -1.8% | -0.02 | -5.9% | 6.5% | -1.0% |
| 40% | -0.01 | -5.2% | 3.8% | -2.6% | 0.00 | -1.4% | 7.3% | -0.7% | -0.01 | -5.7% | 9.0% | 0.7% |
| 60% | 0.00 | -6.9% | 7.0% | -1.8% | 0.01 | -1.5% | 8.6% | -0.2% | 0.00 | -6.3% | 13.9% | 3.5% |
| 80% | 0.01 | -8.5% | 12.4% | 0.5% | 0.03 | -1.4% | 13.4% | 1.8% | 0.03 | -8.2% | 23.0% | 9.7% |
| 100% | 0.16 | 32.0% | 50.1% | 13.3% | 0.07 | 11.3% | 29.9% | 6.4% | 0.10 | 21.8% | 44.1% | 23.2% |

**Fixed prediction considering the rate of Type 1 overlooking**: As explained in Section 2, the probability of Type 1 overlooking $n$ depends on how much testing resource is assigned for negative prediction modules. When the probability is low, the fixed prediction method could degrade the accuracy of prediction, because the method increases false-positive prediction.

To avoid the degradation, we propose a new method which quits the fixed prediction when the rate (i.e., probability) of Type 1 overlooking is low according to the following procedure:

1. Do nothing until after $0.5m$ fixed-prediction modules have been tested.
2. After each module is tested, count the number of modules where test results are defective on the fixed-prediction modules.
3. Calculate the probability rate, dividing the count of step 2 by the number of tested fixed-prediction modules.
4. When the probability rate is smaller than 25%, proposed method quit fixed prediction.
5. Back to Step 2.

For instance, assume that the number of test target modules is 100, and the value of $0.5m$ is 5 in Fig. 4. Five modules have been tested, and there is one module whose test result is defective. Therefore the probability rate is 20%, and the proposed method quits fixed prediction.

## 4. Experiment

**Settings**: In the experiment, we changed probability of Type 1 overlooking from 20% to 100%, by changing the test results of the datasets (see Fig.2) artificially. Similarly, we set the probability of Type 2 overlooking as 20% based on [2]. Note that the lower bound of the probability of Type 1 overlooking is not larger than the probability of Type 2, because assigned testing resources to negative prediction modules is not larger than that of positive prediction.

We evaluated the following prediction models:

- **Reference**: Models where Type 1 and 2 overlooking never occurs.
- **Ordinary**: Models without fixed prediction
- **Fixed prediction**: Models with fixed prediction
- **Proposed method**: Models with proposed method

When evaluating the models, we randomly sorted the order of modules 40 times and calculated the average of the evaluation criteria acquired from the 40 repetitions. This is because the influence of Type 1 overlooking to prediction accuracy is affected the order. For instance, most modules are predicted as positive on the early iterations of online learning, the influence gets smaller (see Section 3).

We randomly selected three datasets (i.e., ant, prop, and synapse) published on PROMISE and D'Ambros et al. [1] repositories to perform our cross-version defect prediction. Each dataset includes 20 independent variables, which include product metrics such as CK metrics.

To predict defective modules, we used logistic regression, a widely used method in defect prediction [1][3]. As a feature selection method, we applied correlation-based feature selection, which is effective when used with logistic regression [3]. We used AUC, precision, recall and F1 score to evaluate the performance of each prediction model.

**Result of ordinary models**: Table 1 shows the evaluation criteria of reference models, and Table 2 shows the difference between reference and ordinary models. Negative values on Table 2 denotes that the accuracy of ordinary models degrades the reference. Except for probability of Type 1 overlooking is 100%, the degradation of AUC, precision and F1 score were moderate. However, the degradation of recall was over 10%, when the probability rate was equal to or larger than 60%. That is, Type 1 overlooking mainly affects recall of prediction models, and this could cause residual defects and degradation of software quality.

**Result of fixed prediction models**: Table 3 shows the difference between ordinary and fixed prediction models. Positive values on table denotes that the accuracy was improved by the fixed prediction model. Overall, recall was improved much. While for the ant dataset, AUC degraded when percentage of Type 1 overlooking was equal to or less than 60%, and precision degraded over 8%.

**Result of proposed models**: Table 4 shows the difference between ordinal and proposed models. Positive values on table denotes that the accuracy was improved by the proposed method. AUC did not degrade when percentage of Type 1 overlooking was larger than 60%, and the extent of the degradation was very small even when the percentage was 20%. Similarly, when the percentage was equal to or larger than 80%, F1 score improved and the degradation was smaller, compared with Table 3. The degradation of precision was smaller than 9%. Compared with Table 3 and 4, negative influences of fixed prediction was suppressed by the proposed method, even when the percentage was 20%.

## 5. Conclusion

This paper focused on software defect prediction models built using online learning. Although the approach is effective, it is affected by overlooking defects. When modules are predicted as "non-defective," fewer test cases are allocated for those modules. Consequently, even when the module is defective, defects can be overlooked during software testing. This overlooking distorts the learning data utilized by online learning.

To mitigate the influence of overlooking, we propose applying a fixed prediction method, which forcibly turns the prediction as "defective" during the initial stage of online learning. However, the method always turns prediction, even when the overlooking seldom occurs. This could degrade the precision of the defect prediction. To address the issue, we propose a new method that discontinues the fixed prediction method when the rate of occurrence of overlooking is low.

In the experiment, we used three datasets, and artificially manipulated the probability of overlooking. The experimental results showed the following:

- When defect prediction models were built using online learning without the fixed prediction method (i.e., the existing approach), recall was degraded over 10%, when the probability of overlooking was 60% or greater.
- When the models were built using the fixed prediction method, recall of the models was improved significantly. While on two out of three datasets, precision was degraded over 5%, regardless of the probability of the overlooking.
- When the models were built using the proposed method, the AUC and F1 score improved when the probability of overlooking was 80% or greater. Compared with the fixed prediction method, the degradation of AUC, precision, and F1 score was smaller, but improvement of recall from the existing approach was steady.

The result suggests that using the proposed method, even if testing resources are drastically reduced for modules that are predicted as defective, the accuracy and recall of the prediction models are not significantly affected.

## Acknowledgments

This research is partially supported by the Japan Society for the Promotion of Science [Grants-in-Aid for Scientific Research (C) (No.21K11840).